\documentclass[a4paper,onecolumn,published,12pt, accepted=2023-06-21]{quantumarticle}
\pdfoutput=1
\usepackage[utf8]{inputenc}
\usepackage[english]{babel}
\usepackage[T1]{fontenc}
\usepackage{tikz}
\usepackage{pgfplots}
\usepackage{cancel}

\usepackage{fullpage}
\usepackage{mdframed}
\usepackage{amsmath, amssymb, amsthm,verbatim, graphicx, xcolor,bbm}
\usepackage{stmaryrd} 
\usepackage{url}

\usepackage{mathrsfs} 

\usepackage{braket}

\makeatletter
\newtheorem*{rep@theorem}{\rep@title}
\newcommand{\newreptheorem}[2]{%
\newenvironment{rep#1}[1]{%
 \def\rep@title{#2 \ref{##1}}%
 \begin{rep@theorem}}%
 {\end{rep@theorem}}}
\makeatother

\newreptheorem{theo}{Theorem}

\DeclareUnicodeCharacter{00A0}{ }

\def\tr{\mbox{tr}}

\def\a{\hat{a}}
\def\bb{\hat{b}}

\def\1{\mathbbm{1}}

\definecolor{colour-for-al}{rgb}{1,0.2,0}

\usepackage{hyperref}
\usepackage[numbers,sort&compress]{natbib}

\usepackage{physics}
\allowdisplaybreaks

\usepackage{graphicx}
\usepackage{caption}
\usepackage{subcaption}
\usepackage{wrapfig}
\usepackage{float}

\begin{document}

\title{The $2T$-qutrit, a two-mode bosonic qutrit}
\author{Aurélie Denys}
\email{aurelie.denys@inria.fr}
\author{Anthony Leverrier}
\orcid{0000-0002-6707-1458}
\email{anthony.leverrier@inria.fr}
\affiliation{Inria Paris, France}

\maketitle

\begin{abstract}
Quantum computers often manipulate physical qubits encoded on two-level quantum systems. Bosonic qubit codes depart from this idea by encoding information in a well-chosen subspace of an infinite-dimensional Fock space. This larger physical space provides a natural protection against experimental imperfections and allows bosonic codes to circumvent no-go results that apply to states constrained by a 2-dimensional Hilbert space. 
A bosonic qubit is usually defined in a single bosonic mode but it makes sense to look for multimode versions that could exhibit better performance. 

In this work, building on the observation that the cat code lives in the span of coherent states indexed by a finite subgroup of the complex numbers, we consider a two-mode generalisation living in the span of 24 coherent states indexed by the binary tetrahedral group $2T$ of the quaternions. The resulting $2T$-qutrit naturally inherits the algebraic properties of the group $2T$ and appears to be quite robust in the low-loss regime. We initiate its study and identify stabilisers as well as some logical operators for this bosonic code. 
\end{abstract}

\section{Introduction}
Quantum information is exquisitely fragile. Quantum circuits are prone to errors and the number of useful gates that can be performed on a quantum computer is thus currently limited to values orders of magnitude away from what is needed for interesting applications. To overcome this problem, two main solutions are possible: better hardware (that is, better \emph{physical} qubits), and quantum error correction (better \emph{logical} qubits). Both approaches can be pursued in parallel, but recent work has shown that it is fruitful to combine them. This is for instance the idea behind \emph{bosonic codes} where a qubit (or a qudit) is encoded into a quantum harmonic oscillator -- a bosonic mode -- of a bosonic system~\cite{BL05,WPG12,ser17}.

While bosonic codes have been studied for a very long time~\cite{CY95,CLY97}, notably in the context of quantum communication~\cite{NAC08,BL16} and optical quantum computing~\cite{KLM01,RGM03,RHG05,WB07}, the GKP qubits~\cite{GKP01} and the cat qubits~\cite{CMM99,MLA14} have known a renewed surge of interest recently as hardware-efficient approaches to fault-tolerant quantum computing with superconducting implementations~\cite{TCV20,GP21}.
An important advantage of bosonic codes is that they considerably simplify the relevant error model and in first approximation, one can benchmark different code families by studying their performance against the pure-loss bosonic channel~\cite{AND18}. These codes exploit the redundancy offered by the infinite dimensionality of the ambient Fock space to give qubits naturally protected against photon loss~\cite{MSB16}. 
These good qubits can then be further protected with code concatenation, for instance with the surface code~\cite{VAW19,CNA22} or even simply with a classical repetition code~\cite{GM19}. 

A natural improvement over these concatenated schemes would be to rely directly on \emph{multimode bosonic codes}. So far, the most studied multimode code family is certainly that of multimode GKP, which can be easily defined by considering higher-dimensional lattices instead of 2-dimensional lattices in phase space for single-mode codes~\cite{GKP01, CEA22}. Generalising other bosonic codes is more complex because one might lose some structure when increasing the dimension and numerical simulations quickly become intractable as we increase the number of modes. As a notable exception, pair-cat codes defined in 2 bosonic modes are particularly robust against phase drift in both modes and loss in a single mode, and can be generalised to a larger number of modes~\cite{AMG19}. Other multimode codes have been considered in the literature~\cite{BL16,NCS18,OC20}.

In addition to bosonic qubits, one may consider higher-dimensional systems. 
Although qubits are appreciated as the direct quantum analogues of classical bits, qudits are also relevant for quantum computing~\cite{WHS20}, and can even lead to better performance, for instance in the case of magic state distillation~\cite{CAB12,cam14,KT19}. 
Bosonic codes are particularly well suited for the design of qudits since the underlying Hilbert space (the harmonic oscillators) is not inherently two-dimensional but infinite-dimensional. The GKP qubits and cat qubits both admit qudit generalisations for instance~\cite{GKP01,HP01,BL16oct}.\\

In this paper, guided by the fact that both the cat qubit and the GKP qubit are naturally obtained as superpositions of structured sets\footnote{The corresponding constellations of states are discrete subgroups of $\mathbbm{C}$.} of coherent states, we consider bosonic codes that live in the span of a finite number $m$ of possibly multimode coherent states. Such a constellation will be of the form $\mathcal{C} = \{|\alpha_1\rangle , \ldots, |\alpha_m\rangle \}$ with $m$ vectors $\alpha_i \in\mathbbm{C}^n$ that correspond to $n$-mode coherent states. The main advantage of this approach\footnote{We note that considering finite constellations of coherent states is not a new idea: see for instance the \emph{group codes for the Gaussian channel} as introduced by Slepian~\cite{sle68},~\cite{Loe91}, as well as their quantum version~\cite{LRS17}.} is that it significantly simplifies the study of the pure-loss channel, which can be represented exactly with a set of $m$ Kraus operators, without any need for truncation. Any such simplification is most welcome since looking for good qudits in an infinite-dimensional space is akin to looking for a needle in a haystack. 
We exploit this formalism to study a nicely symmetric constellation in 4 dimensions (\textit{i.e.}~two bosonic modes): the 24-cell, whose vertices correspond to a finite subgroup of the units of the quaternions, called the \emph{binary tetrahedral group} ($2T$-group). This is reminiscent of the cat code which is naturally defined on a finite subgroup of the units of the complex numbers, and more generally, of rotation-symmetric codes~\cite{GCB20}. 
Once we have defined this 24-dimensional space, we still want to define a logical qudit of smaller dimension. It turns out that this space hosts a very natural qutrit, which we call the $2T$-qutrit, and that we will study in more detail here. We will further see that it is possible to define a $2T$-qubit for which the Pauli-$X$ gate and the phase-gate $P_{2\pi/3} = \left[ \begin{smallmatrix} 1 & 0 \\ 0 & e^{2\pi i/3}\end{smallmatrix} \right]$ can be implemented as passive Gaussian transformations.  \\

The main contribution of this work is to open a new path to explore multi-mode bosonic codes that generalise cat codes and share many important features with them. The structure of single-mode cat codes is related to the group structure of finite subgroups of $U(1)$. Likewise, we present a two-mode code that exploits the properties of the binary tetrahedral group $2T$, which admits a two-dimensional representation in $U(2)$. These ideas can be further generalised, and have inspired a later work defining \textit{quantum spherical codes}, built from arbitrary real or complex polytopes~\cite{JIB23}.
Beyond the theoretical appeal of such generalisations, we also expect that many of the experimental techniques developed to implement single-mode cat codes can be leveraged to prepare and manipulate multimode bosonic systems, such as quantum states from the $2T$-qutrit.\\

The outline of the paper is as follows. We first explain our choice of constellation in Section \ref{sec:constellation}, namely the smallest constellation with a multiplicative group structure that fully exploits the whole phase space of 2 bosonic modes. The next step is to find an interesting subspace that will define a bosonic code. 
To do so, we take inspiration from the cat codes whose logical spaces can be defined from coset states of subgroups of the cyclic group. We define the $2T$-qutrit similarly by considering the 3 cosets of the quaternion group, which is a subgroup of $2T$ with 8 elements. We initiate its study in Section \ref{sec:2T}, and report in Section \ref{sec:simul} our results from numerical simulations assessing its performance against both photon loss and dephasing.

\section{Choice of the constellation}
\label{sec:constellation}

The two most popular bosonic codes consist of superpositions of coherent states from a well-chosen constellation of coherent states in phase space, \textit{i.e.}\ a well-chosen set of numbers in the complex plane. In the case of the (square) GKP code, the constellation is made of (a scaled version of) Gaussian integers of the form $a + bi$ for integers $a, b$ (see Appendix $D$ of~\cite{AND18} for explicit expressions) thus forming an infinite-dimensional grid in the complex plane. For cat codes with $m$ components the constellation consists in the $m$ roots of unity and thus lies on a circle. 
These two sets correspond respectively to additive and multiplicative groups of the complex numbers. 
This is a very useful property since \emph{in fine} the goal is to perform logical operations on the bosonic qubits (or qudits), and some of these operations may correspond to group operations on the constellation. For instance, a logical Pauli-$X$ can be implemented by a translation in phase space (\textit{i.e.}\ the addition of a group element) for the GKP qubit and by a phase-shift (\textit{i.e.}\ a multiplication) for the cat qubit~\cite{TMA22}.
Generalising to $n>1$ modes, it is easy to define new codes with an additive group structure: these are the multimode GKP codes that have been well-studied in the literature~\cite{GKP01,CEA22}. 
Defining codes with a multiplicative group structure seems less well understood. 

Recall that for a single mode, a constellation with a multiplicative group structure is necessarily a finite subgroup of the complex units, namely the set $\{ z^k \: : \: 0 \leq k \leq m-1\}$ where $z = e^{2\pi i/m}$ is an $m$-root of unity. If we pick an amplitude $\alpha >0$, such a constellation defines a set of $m$ coherent states through the obvious identification of phase space with the complex plane, $\{ |z^k \alpha\rangle \: : \: 0 \leq k \leq m-1\}$,
where we recall that the coherent state $|\gamma\rangle$ is an eigenvector of the annihilation operator $\hat{a}$ with eigenvalue $\gamma \in \mathbbm{C}$. Its expansion in the Fock basis is given by $|\gamma \rangle = e^{-|\gamma|^2/2} \sum_{n=0} ^\infty\frac{\gamma^n}{\sqrt{n!}} |n\rangle$, where $|n\rangle$ is the Fock state with $n$ photons.

We wish to define a multimode version of these cat codes by considering the simplest possible instance, namely the 2-mode case. 
When moving to 2 modes, the phase space becomes 4-dimensional and can be identified with the division algebra of quaternions $\mathbbm{H}$, that is the numbers of the form $a + bi + cj + dk$, where $a, b, c, d$ are arbitrary real numbers and $i,j,k$ satisfy
\begin{equation}
    i^2 = j^2 = k^2 = ijk = -1.
\end{equation}

It will somewhat simplify the notations later on to introduce a parameter $\beta = \alpha (1+i)$ and to associate the two-mode coherent state to a quaternion through the identification:
\begin{equation}
    a + bi + cj + dk \in \mathbbm{H} \qquad \mapsto \qquad |(a+bi) \beta\rangle |(c-di)\beta\rangle \in \mathrm{Span}(\lbrace|n_1, n_2\rangle \: : \: n_1, n_2 \in \mathbbm{N}\rbrace).
\end{equation}

Here $|n_1,n_2\rangle$ denotes a Fock space with $n_1$ photons in the first mode and $n_2$ photons in the second mode.

Inspired by the single-mode case, we want a constellation corresponding to a multiplicative subgroup of the quaternions. 
The finite subgroups have been classified~\cite{cox91}: 
\begin{enumerate}
\item the cyclic groups of order $m$, for $m \in \mathbbm{N}$,
\item the dicyclic groups of order $4p$, for $p \in \mathbbm{N}$,
\item the binary tetrahedral group, denoted $2T$, of order 24,
\item the binary octahedral group, of order 48,
\item the binary icosahedral group, of order 120.
\end{enumerate}
The cyclic groups simply give the single-mode cat states, so do not yield genuine 2-mode bosonic codes. The dicyclic groups give constellations of the form $\{ | e^{i \pi k/p} \alpha\rangle |0\rangle, |0\rangle | e^{i \pi \ell/p}\rangle \: : \: 0 \leq k,\ell \leq 2p-1\}$. The states then correspond to the superposition of a cat state in one mode and the vacuum state in the second mode. The three remaining subgroups look more intriguing since they cannot be directly obtained from subgroups of the unit complex numbers. 
Given that both the binary octahedral and binary icosahedral groups are quite large, we choose to focus here on the binary tetrahedral group, which already promises to pose significant challenges for implementation! We also remark that the elements of the group $2T$ form the vertices of the 24-cell, one of the rare regular polytopes in 4 dimensions. This generalises the single-mode case where 2-dimensional regular polygons are naturally associated with the $m$-roots of unity. Intuitively, this property should correlate with a better resistance of the bosonic code against dephasing. 

The binary tetrahedral group $2T$ is the following set of 24 quaternions:
\begin{equation}
    \big\{ \pm 1, \pm i, \pm j, \pm k, \frac{1}{2}( \pm 1 \pm i \pm j \pm k)\big\}, \label{eq: def 2T}
\end{equation}
with all possible sign combinations.
The 24 coherent states in the constellation define a 24-dimensional Hilbert space
\begin{equation}
    \mathcal{H}_{2T} := \mathrm{Span}( \{|i^k \beta\rangle |0\rangle, |0\rangle |i^\ell \beta\rangle, |e^{i k\pi/2} \alpha\rangle |e^{i \ell \pi/2} \alpha\rangle \: : \: 0\leq k, \ell \leq 3\})
\end{equation}
where $\alpha >0$ is arbitrary and we defined $\beta = \alpha (1+i)$. In practice, there will exist some optimal values of $\alpha$, known as sweet spots for cat codes~\cite{AND18}, unless one is interested in biased-noise qubits, in which case larger values of $\alpha$ are typically preferred~\cite{GM19,PSG20,CNA22}. \\

This choice of constellation in phase space naturally defines a 24-dimensional space. Arguably, this remains quite a large dimension, and our next goal will be to define a qudit of smaller dimension, namely a qutrit, within this space. 
Before we do that, let us comment on the possibility of stabilising the 24-dimensional space corresponding to $\mathcal{H}_{2T}$.

\paragraph{Jump operators for $\mathcal{H}_{2T}$.} 
Some bosonic codes such as the cat code and the pair-cat code~\cite{AMG19} can be realised by reservoir-engineering. This is for instance the case of the four-component cat and pair-cat codes. Defining the dissipator $\mathcal{D}[F](\rho) = F\rho F^\dag - \frac{1}{2}\{F^\dag F, \rho\}$ for a jump operator $F$ and a state $\rho$, they respectively have for jump operators 
\begin{equation}
    F_{\mathrm{cat}} = \a^4 - \alpha^4, \qquad \text{ and } \qquad F_{\text{pair-cat}} = \a^2 \bb^2 - \gamma^4,
\end{equation}
where the parameters $\alpha, \gamma$ control the amplitudes of the codes.
In particular, any code state is annihilated by the corresponding jump operator and is a fixed point of the time evolution $\dot{\rho} = \kappa \mathcal{D}[F](\rho)$, for any value of the parameter $\kappa >0$. 

Similarly to these two codes, it is possible to find jump operators for the 24-dimensional Hilbert space $\mathcal{H}_{2T}$, for instance, 
\begin{align}
F_1 &= (\a^4 + \hat{b}^4 + \alpha^4)^2 - 9 \alpha^8, \label{eqn:F1}\\
F_2 &= 6 \a^4 \hat{b}^4 - \alpha^4(\a^4+\hat{b}^4) - 4 \alpha^8. \label{eqn:F2}
\end{align}
To see this, let us determine which 2-mode coherent states $|\beta_1\rangle |\beta_2\rangle$ satisfy
\begin{equation}
    F_1 |\beta_1\rangle |\beta_2\rangle = F_2|\beta_1\rangle |\beta_2\rangle=0.
\end{equation}
The first equation immediately gives $\beta_1^4 + \beta_2^4 \in \{ 2 \alpha^4, -4\alpha^4\}$.
Assuming the first case, the second equation implies that $\beta_1^4 \beta_2^4 = \alpha^8$ and therefore $\beta_1^4=\beta_2^4 = \alpha^4$. This gives the 16 coherent states of the form $|e^{i k\pi/2} \alpha\rangle |e^{i \ell \pi/2} \alpha\rangle $. 
The second case implies that $\beta_1^4 \beta_2^4=0$, and together with $\beta_1^4+\beta_2^4 = -4\alpha^4$, we obtain the 8 states of the form $|i^k  \alpha(1+i)\rangle |0\rangle$ or $|0\rangle |i^\ell  \alpha(1+i)\rangle$.
It is immediate to check that the 24 states of the constellation are indeed annihilated by $F_1$ and $F_2$.
%

\section{The $2T$-qutrit}
\label{sec:2T}
In this section, we introduce the $2T$-qutrit which is a 3-dimensional subspace of $\mathcal{H}_{2T}$ and study some of its properties. In particular we find its stabilisers as well as some logical operators. 

\subsection{Definition of the $2T$-qutrit}
To define the $2T$-qutrit, we will take inspiration from the cat codes. Let us first recall that one can define a bosonic cat code with $m$ components by considering a specific subspace of $\mathrm{Span}( \{ |z^k \alpha\rangle \: : \: 0 \leq k \leq m-1\})$ where $z := e^{2\pi i/m}$ is an $m$-root of unity. 
More precisely, if $m = dn$, one can define a qudit of dimension $d$ by considering the states
\begin{equation}
    |\chi_k\rangle \propto \sum_{\ell = 0}^{n-1} |z^{\ell n + k} \alpha \rangle = \sum_{u \in z^k \mathbb{U}_n} | u\rangle,
\end{equation}
if we formally associate the coherent state $|u\rangle := |(a +bi) \alpha\rangle$ to the complex number $u = a+ bi \in \mathbbm{C}$. Here, $\mathbb{U}_n$ is the group of $n$-roots of unity.
The states $|\chi_k\rangle$ therefore correspond to the $d$ cosets of $\mathbb{U}_n$ in $\mathbb{U}_{dn}$.
An immediate property of these states is that a phase-shift $P := e^{2\pi \hat{a}^\dag \hat{a}/(dn)}$ will map $|z^j\rangle$ to $|z^{j+1}\rangle$ and therefore $|\chi_k\rangle$ to $|\chi_{k+1}\rangle$, where the indices are understood modulo $d$. 
The states $|k_L\rangle := \sum_{\ell=0}^{n-1} e^{\frac{-2\pi i k\ell}{d}} |\chi_\ell\rangle$ are eigenstates of $P$ with eigenvalue $e^{2\pi ki/d}$ and form an orthonormal basis of the space. In particular, the operator $P$ acts as a logical Pauli-$Z$ on this qudit.

We remark that this method can be used to construct a qudit of dimension $d$ from a group $G$ whenever there exists a subgroup $H$ of $G$ with cosets of the form $g^k H$ for one $g \in G$ and $g^d H = H$. 
The binary tetrahedral group $2T$ (eq:\ref{eq: def 2T}) can be obtained as the semi-direct product of the quaternion group $Q = \{ \pm 1, \pm i, \pm j, \pm k\}$ with the cyclic group $C_3 = \{1,\omega, \omega^2\}$ generated by the element $\omega = -\frac{1}{2} (1+i+j+k)$:
\begin{equation}
    2T= Q \rtimes C_3.
\end{equation}
It thus satisfies the above conditions for $H=Q$, $g= \omega$ and $d=3$. We exploit this decomposition to define our qutrit.

Let us therefore introduce the three states:
\begin{align}
|\phi_0\rangle := \nu \sum_{q\in Q} |q\rangle, \quad |\phi_1\rangle := \nu \sum_{q\in \omega Q} |q\rangle, \quad |\phi_2\rangle := \nu \sum_{q\in \omega^2 Q} |q\rangle,
\end{align}
where $\nu = \frac{e^{\alpha^2} }{4 \sqrt{2 + \cos(2\alpha^2) + \cosh(2\alpha^2)}} $
is a normalisation coefficient (see Appendix \ref{app:B} for its derivation) and we recall that we write $|a+bi+cj+dk\rangle$ to mean the 2-mode coherent state $|(a+bi) \beta\rangle |(c-di) \beta \rangle$, where we set $\beta := \alpha (1+i)$. 
The sets $\omega Q := \{\omega q \: : \: q\in Q\}$ and $\omega^2 Q := \{\omega^2 q \: :\: q \in Q\}$ are given by
\begin{align}
\omega Q &= \left\{ \pm \frac{1}{2}(1\!+\!i\!+\!j\!+\!k), \pm \frac{1}{2}(1\! +\! i\!-\! j\!-\! k), \pm \frac{1}{2}(1\!-\!i\!-\!j\!+\!k), \pm \frac{1}{2}(1\!-\!i\!+\!j\!-\!k) \right\},\\
\omega^2 Q &= \left\{ \pm \frac{1}{2}(-1\!+\!i\!+\!j\!+\!k) ,\pm \frac{1}{2}(1\! -\!i\!+\!j\!+\!k), \pm \frac{1}{2}(1\!+\!i\!-\!j\!+\!k), \pm \frac{1}{2}(1\!+\!i\!+\!j\!-\!k)\right\}.
\end{align}
In particular, the set $\omega Q$ contains quaternions with an even number of minus signs, while $\omega^2 Q$ contains those with an odd number of minus signs. 
We define the $2T$-qutrit as $\mathrm{Span}(\{ |\phi_0\rangle, |\phi_1\rangle, |\phi_2\rangle\})$.\\

The states $|\phi_k\rangle$ can be conveniently expressed using single-mode cat states. If we denote these single-mode cat states with 2 or 4 coherent states as 
\begin{align}
|\alpha_2\rangle & := c_2^\alpha (|\alpha\rangle + |-\alpha\rangle),\\
|i\alpha_2\rangle & := c_2^\alpha (|i\alpha\rangle + |-i\alpha\rangle),\\
|\alpha_4\rangle & := c_4^\alpha  ( |\alpha\rangle + |i\alpha\rangle + |-\alpha\rangle + |-i\alpha\rangle),
\end{align}
with normalisation coefficients given by
\begin{align}
c_2^\alpha = \frac{1}{\sqrt{2(1+e^{-2|\alpha|^2})}}, \qquad
c_4^\alpha = \frac{1}{\sqrt{8 e^{-|\alpha|^2} (\cosh |\alpha|^2 + \cos|\alpha|^2)}},
\end{align}
then we obtain that 
\begin{align}
|\phi_0\rangle &\propto |\beta_4\rangle|0\rangle + |0\rangle |\beta_4\rangle \label{eqn:phi0},\\
|\phi_1\rangle & \propto |\alpha_2\rangle |i \alpha_2\rangle + |i \alpha_2\rangle |\alpha_2\rangle, \label{eqn:phi1}\\
|\phi_2\rangle &\propto |\alpha_2\rangle |\alpha_2\rangle + |i \alpha_2\rangle |i \alpha_2\rangle.\label{eqn:phi2}
\end{align}

While the three states above are not orthogonal, one finds an orthonormal basis of the qutrit by defining:
\begin{align} \label{basis}
 |\bar{k}\rangle := \nu_k \sum_{\ell=0}^2 \zeta^{-k\ell} |\phi_\ell\rangle, \qquad \text{for} \, k \in \{0,1,2\}
\end{align}
where $\zeta = e^{\frac{2 \pi i }{3}}$ is a cubic-root of unity, and where the normalisation coefficients $\nu_0$ and $\nu_1=\nu_2$ are given in \eqref{eqn:nu0} and \eqref{eqn:nu12} in Appendix \ref{app:B}.
To show that the basis $\{ |\bar{k}\rangle\}$ is indeed orthogonal, we will introduce a unitary operator that acts as the logical Pauli-$Z$ operator for the qutrit, namely $\bar{Z} |\bar{k}\rangle = \zeta^k |\bar{k}\rangle$.

Let us first define the unitary matrix 
\begin{equation}
    U: = -\frac{1}{2} \begin{bmatrix} 1+i & -1-i\\ 1-i & 1-i\end{bmatrix} = \frac{1}{\sqrt{2}} \begin{bmatrix} e^{-3i\pi/4} & e^{i\pi/4}\\ e^{3i\pi/4}& e^{3i\pi/4}\end{bmatrix}.
\end{equation}
It corresponds to the representation of the quaternion $\omega$ via the map
\begin{equation} \label{eqn:repr}
  \rho : a+bi+cj+dk \in \mathbb{H} \mapsto 
  \begin{bmatrix}
  a+bi & -c-di \\
  c-di & a-bi
  \end{bmatrix} \in \mathcal{M}_2(\mathbb{C}).
\end{equation}
In particular, one can check that $U^3 = (\rho(\omega))^3 = \1_2$.
Moreover, if we denote by $\mathcal{U}$ the operator acting on the two-mode Fock space by mapping the two-mode coherent state $ |\alpha_1\rangle |\alpha_2\rangle$ to $|\alpha'_1\rangle |\alpha'_2\rangle$ defined as $ \left[\begin{smallmatrix}
  \alpha'_1\\ \alpha'_2
  \end{smallmatrix}\right] = U \left[\begin{smallmatrix}
  \alpha_1\\ \alpha_2
  \end{smallmatrix}\right]$, then we have $\mathcal{U}|q\rangle = | \omega q\rangle$ for any quaternion $q \in 2T$, and therefore
  \begin{equation}
      \mathcal{U}|\phi_\ell\rangle = |\phi_{\ell+1}\rangle
  \end{equation}
by definition of the sets $Q$, $\omega Q$ and $\omega^2 Q$. 
Exploiting \eqref{basis}, we obtain 
\begin{align}
\mathcal{U}|\bar{k}\rangle &= \nu_k \sum_{\ell=0}^2 \zeta^{-k\ell} |\phi_{\ell+1}\rangle = \nu_k \zeta^k \sum_{\ell=0}^2 \zeta^{-k\ell} |\phi_{\ell}\rangle = \zeta^k|\bar{k}\rangle,
\end{align} 
where the indices are always understood modulo 3. 
This shows that $\mathcal{U}|\bar{k}\rangle = \zeta^k |\bar{k}\rangle$, implying that the three states are eigenstates of the unitary $\mathcal{U} = \bar{Z}$ with distinct eigenvalues, and are therefore orthogonal. 
One can also write the expression of $\mathcal{U}$ as a Gaussian passive transformation~\cite{leo03}:
\begin{align}\label{eqn:Z}
 \bar{Z} = \exp \Big(\frac{2\pi}{3\sqrt{3}} i (-a^\dag a +(1-i) a^\dag b + (1+i) ab^\dag +b^\dag b)\Big).
 \end{align}

Before studying the $2T$-qutrit in more detail, it is instructive to compute the limit of the logical states when $\alpha \to 0$. We find that, up to unessential global phases,
\begin{align}
|\bar{0}\rangle & \xrightarrow[\alpha \, \to \,  0]{} |00\rangle,\\
|\bar{1}\rangle & \xrightarrow[\alpha \, \to \, 0]{} \frac{1}{2} (|40\rangle + |04\rangle) - \frac{i}{\sqrt{2}} |22\rangle,\\
|\bar{2}\rangle & \xrightarrow[\alpha\, \to \, 0]{} \frac{1}{2} (|40\rangle + |04\rangle) + \frac{i}{\sqrt{2}} |22\rangle.
\end{align}
In this limit, the states $\frac{1}{\sqrt{2}} ( |\bar{1}\rangle \pm |\bar{2}\rangle )$ take the simple expressions $\frac{1}{\sqrt{2}} (|40\rangle + |04\rangle)$ and $ |22\rangle$ which coincide with an instance of the Chuang-Leung-Yamamoto code~\cite{CLY97}.
We will discuss the bosonic qubit $\mathrm{Span}(\{|\bar{1}\rangle, |\bar{2}\rangle\})$ in more detail in Section \ref{sec:qubit}.

\subsection{Stabilisers}

We have seen in Section \ref{sec:constellation} that the operators $F_1+1$ and $F_2+1$ stabilise the 24-dimensional Hilbert space $\mathcal{H}_{2T}$. 
In order to stabilise the $2T$-qutrit, we need additional stabilisers.
Let us denote by $\hat{n}_1 = \hat{a}^\dag \hat{a}$ and $\hat{n}_2 = \hat{b}^\dag \hat{b}$ the photon number operators in the two modes and introduce the phase operators $R_1:= e^{i \hat{n}_1 \pi/2}, R_2:= e^{i \hat{n}_2 \pi/2}$ and the SWAP operator $e^{i (\hat{a}^\dag- \bb^\dag)(\a-\bb)\pi/2}$ that exchanges the two modes. 

Recalling how the phase operators act on cat states, 
\begin{equation}
    e^{i \hat{n} \pi/2} |\alpha_2\rangle = |i\alpha_2\rangle, \qquad e^{i \hat{n} \pi/2} |i \alpha_2\rangle = |\alpha_2\rangle, \qquad e^{i \hat{n} \pi/2} |\beta_4\rangle = |\beta_4\rangle  ,
\end{equation}
it is immediate from \eqref{eqn:phi0}, \eqref{eqn:phi1}, \eqref{eqn:phi2} that $R_1 R_2$ and $R_1^2$ and the SWAP operator stabilise the $2T$-qutrit since they leave the states $|\phi_k\rangle$ invariant. 
One can also check that the only states of $\mathcal{H}_{2T}$ stabilised by $R_1 R_2, R_1^2$ and $\mathrm{SWAP}$ are states of the $2T$-qutrit.

To summarise, the $2T$-qutrit is exactly the set of states stabilised by $F_1+1, F_2+1, R_1 R_2, R_1^2, \text{ and the } \mathrm{SWAP}$:
\begin{equation}
    \mathrm{Span}(\{ |\bar{0}\rangle, |\bar{1}\rangle, |\bar{2}\rangle\}) = \Big\{ |\psi\rangle \: \text{s.t.} \: S |\psi\rangle = |\psi\rangle \quad \forall S \in \mathcal{S} \Big\},
\end{equation}
where we define the set of stabilisers as 
\begin{equation}
    \mathcal{S} =\{ F_1+1, F_2+1, e^{i (\hat{n}_1+\hat{n}_2) \pi/2}, e^{i \hat{n}_1 \pi}, e^{i (\hat{a}^\dag- \bb^\dag)(\a-\bb)\pi/2}\}.
\end{equation}

A simple consequence is the existence of invariants for the states in the $2T$-qutrit. In particular, the photon numbers $n_1, n_2$ in both modes are restricted to specific values:
\begin{equation}
    n_1+ n_2 \equiv 0 \mod 4, \quad n_1 \equiv 0 \mod 2, \quad n_2 \equiv 0 \mod 2.
\end{equation}

\subsection{A remarkable $2T$-qubit within the $2T$-qutrit}
\label{sec:qubit}

We observe that 
\begin{equation}
    R_1 |\phi_0\rangle = |\phi_0\rangle, \quad R_1 |\phi_1\rangle = |\phi_2\rangle, \quad R_1 |\phi_2\rangle = |\phi_1\rangle,
\end{equation}
since $R_1$ leaves the cat state $|\beta_4\rangle$ invariant and exchanges $|\alpha_2\rangle$ and $|i \alpha_2\rangle$.
Said otherwise, $R_1 |\phi_\ell\rangle = |\phi_{2\ell}\rangle$, with the index understood modulo 3. The operator therefore acts as follows on the logical states
\begin{align}
R_1 |\bar{k}\rangle = \nu_k \sum_{\ell=0}^2 \zeta^{-k\ell} R_1 |\phi_\ell\rangle = \nu_k \sum_{\ell=0}^2 \zeta^{-k\ell} |\phi_{2\ell}\rangle =\nu_k\sum_{\ell=0}^2 \zeta^{-2k\ell} |\phi_{\ell}\rangle = |\overline{2k}\rangle,
\end{align}
since $\nu_k = \nu_{2k}$.

It means that $R_1$ and $R_2$ act as a gate $X_{12}$ on the $2T$-qutrit, with 
\begin{equation}
    X_{12} = \left[\begin{matrix} 1 & 0 & 0\\ 0 &  0 & 1 \\ 0 &1& 0 \end{matrix}\right].
\end{equation}

Interestingly, this means that if we restrict ourselves to the qubit space $\mathrm{Span}(\{|1\rangle, |2\rangle\})$, then there exist two Gaussian passive transformations acting as
\begin{equation}
    \mathcal{U} = \zeta \begin{bmatrix} 1 & 0 \\ 0 & \zeta \end{bmatrix}, \qquad \text{ and } \qquad R_1 = \begin{bmatrix} 0 & 1 \\ 1 & 0\end{bmatrix}.
\end{equation}
In other words, $R_1$ acts on the qubit as a logical Pauli-$X$ gate while $\mathcal{U}$ acts as a logical phase gate $P(2\pi/3)$ of angle $2\pi/3$.\\

A state of the $2T$-qubit takes a particularly simple form:
\begin{align}
\frac{1}{\sqrt{2}} (|\bar{1}\rangle - |\bar{2}\rangle) &= \frac{\nu_1}{\sqrt{2}} \sum_{\ell=0}^2 (\zeta^{-\ell} - \zeta^{-2\ell})|\phi_\ell\rangle\\
&\propto |\phi_1\rangle - |\phi_2\rangle\\
&\propto  |\alpha_2\rangle |i \alpha_2\rangle + |i \alpha_2\rangle |\alpha_2\rangle -  |\alpha_2\rangle |\alpha_2\rangle - |i \alpha_2\rangle |i \alpha_2\rangle & \text{from} \; \eqref{eqn:phi1} \; \text{and} \; \eqref{eqn:phi2} \\
&\propto ( |\alpha_2\rangle - |i \alpha_2\rangle)^{\otimes 2},
\end{align}
which corresponds to a product state of two four-component cat qubit states. 
Admittedly, recent experimental progress on the single-mode cat qubits indicates that preparing the state $\frac{1}{\sqrt{2}} (|\bar{1}\rangle - |\bar{2}\rangle)$ should not be completely out of reach.


\section{Numerical simulations}
\label{sec:simul}

In this section, we study the performance of the $2T$-qutrit against either pure loss or dephasing. We note that some recent studies~\cite{LXJ22} consider both sources of noise at the same time, but only for single-mode codes. Because the $2T$-qutrit is a two-mode code, the relevant subspace of the Fock space required to perform accurate simulations quickly becomes very large even for moderate values of $\alpha$. For instance, truncating to Fock states $|n_1, n_2\rangle$ with $n_1+n_2 \leq N$ yields a Hilbert space of dimension $(N+1)(N+2)/2$. 
On the other hand, the specific structure of the $2T$-qutrit leads to significant simplifications when the channel is either a pure-loss or a dephasing channel. 
In the first case, we can exploit the fact that the $2T$-qutrit is defined as a subspace of the 24-dimensional space spanned by 2-mode coherent states. For the dephasing channel, we exploit the fact that the photon number is invariant under dephasing, and therefore that it is possible to represent the relevant states in a compact form since the Fock states $|n_1, n_2\rangle$ necessarily satisfy $n_1 +n_2 \equiv 0 \mod 4$ and $n_1 \equiv 0 \mod 2$. 

Throughout this section, we assume that each mode of the $2T$-qutrit is affected independently by the same quantum channel. In particular, if the single-mode channel describing pure loss or dephasing is denoted by $\mathcal{N}$, then the overall two-mode quantum channel is given by $\mathcal{N} \otimes \mathcal{N}$. While this independence seems like a reasonable assumption for losses, it may be too pessimistic in the case of dephasing. \\

The python files used to perform the numerical analysis are available on \href{https://gitlab.inria.fr/adenys/the-2t-qutrit}{Gitlab}.

\subsection{A relevant figure of merit: the entanglement fidelity}
\label{sec:optim}

In this subsection, we consider the general case of an encoding in the span of $m$ potentially multimode coherent states $|\alpha_1\rangle, \ldots, |\alpha_m\rangle$. We will instantiate it with the set $2T$ in Sections \ref{sub:loss} and \ref{sub:dephasing}.

Since the goal is to measure how ``close'' the recovered state is from the original one, it is natural to choose a metric based on the fidelity\footnote{There are of course many other reasonable metrics one could consider, for instance the maximum number of photon losses or additions that can be corrected perfectly, as is done in~\cite{JIB23}.}. In particular, \cite{FSW07} considers three common options and argues that the entanglement fidelity is a convenient choice because maximising over the possible recovery operations then simply amounts to solving a semi-definite program. As we will see, this is also the case if one aims at optimising the encoding for a given recovery operation. The entanglement fidelity measures how well a channel preserves the entanglement. It is related to the average input-output fidelity of a channel and also has other useful properties reviewed in Appendix A of \cite{AND18}.
It is defined as
\begin{align}
F(\mathcal{C}) := \langle \Phi_d | \mathrm{id} \otimes \mathcal{C} (|\Phi_d\rangle \langle \Phi_d|) |\Phi_d\rangle,
\end{align}
where $|\Phi_d\rangle := \frac{1}{\sqrt{d}} \sum_{i=0}^{d-1} |i\rangle |i\rangle \in \mathbbm{C}^d \otimes \mathbbm{C}^d$ is the $d$-dimensional maximally entangled state, $\mathcal{C}$ is the channel under consideration, and $\mathrm{id}$ denotes the identity channel on the first subsystem. Note that this is the fidelity between a maximally entangled state and a maximally entangled state whose second system has been sent through the channel $\mathcal{C}$. Here, the channel of interest will be $\mathcal{C} = \mathcal{E} \circ \mathcal{N} \circ \mathcal{R}$: the second system will be first encoded in the code via $\mathcal{E}$, sent through the noise channel $\mathcal{N}$ (either a pure-loss channel or a pure-dephasing channel) and its output will then be decoded with a recovery operation $\mathcal{R}$. More precisely, we will consider $\mathcal{F}(\mathcal{E}):=\max_{\mathcal{R}} F(\mathcal{E}\circ \mathcal{N} \circ \mathcal{R})$ where one maximises over all recovery maps. Ideally it would be better to consider the best recovery map that can be practically realised for each encoding. However, the notion of ``practically realisable operations'' is not so well defined and taking the optimal map has the benefit of considerably simplifying the problem while retaining the advantage of putting the encodings on an equal footing. This was also the strategy in~\cite{AND18} which benchmarked various single-mode bosonic codes.

A qudit code of dimension $d$ is defined by an encoding map 
\begin{align}
\mathcal{E}: \left\{\begin{array}{ccc}
\mathcal{B}(\mathbbm{C}^{d}) & \to& \mathcal{B}(\mathrm{Span}(\{ |\alpha_k\rangle\}))\\
|k\rangle &\mapsto & |\bar{k}\rangle
\end{array}
\right.
\end{align}
where $\mathcal{B}(\mathcal{H})$ denotes the set of bounded linear operators on the Hilbert space $\mathcal{H}$, and the state $|\bar{k}\rangle$ represents the encoded version of the state $|k\rangle$. The output of the channel $\mathcal{N}$ is a density matrix defined on the two-mode Fock space $\mathfrak{F}(\mathbbm{C}^2) = \mathrm{Span}(|n_1, n_2\rangle \: : \: n_1, n_2 \in \mathbbm{N})$.
In particular, we note the output of the pure-loss channel will belong to a much smaller space, namely the span of the coherent states $|\mu \alpha_k\rangle$, which is finite-dimensional by construction.
We consider a recovery map 
\begin{equation}
    \mathcal{R} : \mathcal{B}(\mathfrak{F}(\mathbbm{C}^2)) \to \mathcal{B}(\mathbbm{C}^{d})
\end{equation}
that describes how the output of the channel is decoded.

To find the best possible encoding we are interested in computing $\max_{\mathcal{E}} \mathcal{F}(\mathcal{E}) = \max_{\mathcal{E},\mathcal{R}} F(\mathcal{E} \circ \mathcal{N} \circ \mathcal{R})$. While the problem of maximising the entanglement fidelity over the choice of $\mathcal{E}$ and $\mathcal{R}$ is typically not amenable to efficient optimisation~\cite{BBF22}, one can proceed as in~\cite{RW05} to find a local optimum by iteratively maximising $F$ while fixing one input (either $\mathcal{E}$ or $\mathcal{R})$ and then fixing the other operator until convergence. The advantage is that both problems $\max_{\mathcal{R}} F(\mathcal{E}\circ \mathcal{N} \circ \mathcal{R})$ and $\max_{\mathcal{E}} F(\mathcal{E}\circ \mathcal{N} \circ  \mathcal{R})$ are semi-definite programs that can be solved efficiently. 

Denoting by $\{E_j\}$, $\{C_k\} $ and $\{R_\ell\}$ the Kraus operators of the encoding map, noise channel and recovery map, we find that the entanglement fidelity is given by
\begin{align}\label{eqn:F}
F (\mathcal{E} \circ \mathcal{N} \circ \mathcal{R}) = \sum_{j,k,\ell} \langle \Phi_d | \1 \otimes R_\ell C_k E_j |\Phi_d \rangle\langle \Phi_d | \1 \otimes E_j^\dag C_k^\dag R_\ell^\dag |\Phi_d\rangle. 
\end{align}
Remark that the two channels $\mathcal{E}$ and $\mathcal{R}$ we want to optimise are also characterised by the positive semi-definite operators $X_E := \sum_j (\1 \otimes E_j) |\Phi_d\rangle \langle \Phi_d| (\1 \otimes E_j^\dag)$ and $X_R := \sum_i (\1 \otimes R_i^\dag) |\Phi_d\rangle \langle \Phi_d| (\1 \otimes R_i)$.
For the initialisation, one can take either a specific encoding (\textit{e.g.}\ that of the $2T$-qutrit) or a random encoding. Then we optimise successively:
\begin{itemize}
\item the recovery map:
\begin{align}\label{eqn:R}
\max_{X_R} \mathrm{tr}(X_R M_E) \quad \text{s.t.} \quad \tr_1 X_R = \frac{1}{d} \1_d,
\end{align}
with $M_E := \sum_k (\1 \otimes C_k) X_E^* (\1 \otimes C_k^\dag)$,
\item and the encoding map:
\begin{align}\label{eqn:E}
\max_{X_E} \mathrm{tr}(X_E N_R) \quad \text{s.t.} \quad \tr_2 X_E = \frac{1}{d} \1_d,
\end{align}
with $N_R := \sum_k (\1 \otimes C_k^\dag) X_R^* (\1 \otimes C_k)$,
\end{itemize}
where $X_E^*$ and $X_R^*$ denote the values of $X_E$ and $X_R$ obtained at the previous step.

This process is not known to converge to the optimal solution in general, but it performs reasonably well provided that the loss parameter $\gamma$ is not too small. In that case, the optimisation tends to consistently converge to an optimum independent of the starting point, suggesting that the corresponding value is in fact close to the global maximum. For instance, starting from a truncated (single-mode) Fock space, this algorithm will converge to an encoding map close to (a displaced version of) the hexagonal GKP code~\cite{NAJ18}.

Our numerical optimisations are realised with the Splitting Conic Solver~\cite{ocpb:16,scs}.

\subsection{Performance for the pure-loss channel}
\label{sub:loss}

One of the advantages of a bosonic encoding is that it greatly simplifies the relevant error model that should be addressed. In particular, as a first approximation, it can be modelled as a pure-loss channel, which is described by an infinite set of Kraus operators
\begin{align}\label{eqn:kraus}
  \set{K_k = c_k \hat{a}^k \mu^{\hat{n}} \: : \: k \in \mathbb{N}}
\end{align}
where $\mu = \sqrt{1-\gamma}$, $c_k = \frac{1}{\sqrt{k!}}(\frac{\gamma}{1-\gamma})^{\frac{k}{2}}$, and $\gamma \in [0,1[$ is the loss parameter. The single-mode loss channel $\mathcal{N}_{L,\gamma}$ thus acts as follows~\cite{AND18}: 
\begin{align} \label{eqn:loss}
\mathcal{N}_{L,\gamma}: \quad \rho \mapsto \sum_{k=0}^\infty K_k \rho K_k^\dag.
\end{align}
When working with the whole Fock space, this representation contains an infinite number of operators and one has to resort to some approximations to perform numerical simulations, for instance a truncation of the Fock space. We avoid this problem since we work instead with a finite-dimensional subspace of the Fock space spanned by $m$ coherent states.
Since the pure-loss channel maps a coherent state to another pure (attenuated) coherent state, we find that this channel can be described by a \emph{finite} set of Kraus operators: see Appendix \ref{sec:kraus} for details. In particular, when working with $m$ possible coherent states, it is possible to find a Kraus representation of the pure-loss channel consisting of $m$ operators. 

Initially, we (naively) wanted to exploit this representation of the pure-loss channel to find good bosonic codes, that is good subspaces of $\mathrm{Span}(\{|\alpha_1\rangle, \ldots, |\alpha_m\rangle\})$. The idea is to compute a figure of merit, for instance the \emph{entanglement fidelity} as defined in section \ref{sec:optim}, and try to find the qubit (or qudit) that maximises this quantity. While this makes sense in theory, it turns out to be very difficult in practice, and the main issue is that the best qubit (according to this figure of merit) will likely be very unstructured, and therefore pretty much useless for understanding how it can be exploited for fault tolerance (which is the long-term objective). 
More realistically, the figure of merit can be used to benchmark the quality of various encodings, by comparing it to the value obtained by numerical optimisation. Moreover, checking that the value obtained for a given encoding corresponds to a local optimum is also an indication that the encoding is not too bad. 

The main advantage of our approach is that the number of Kraus operators of the pure-loss channel is finite. Consequently, the values of the operators $M_E$ and $N_R$ can be computed exactly without resorting to any truncation in the sum, contrary to what is done in~\cite{AND18} for instance. In addition, since there is no need for truncation, it is possible to study the performance even for large values of $\alpha$. This is in contrast with standard methods where larger values of $\alpha$ require to increase the level of truncation to get accurate results. This is extremely useful when considering multimode bosonic codes since even truncated Fock spaces quickly become very large.

\begin{figure}[htbp]
\centering
 \includegraphics[width=0.48\textwidth]{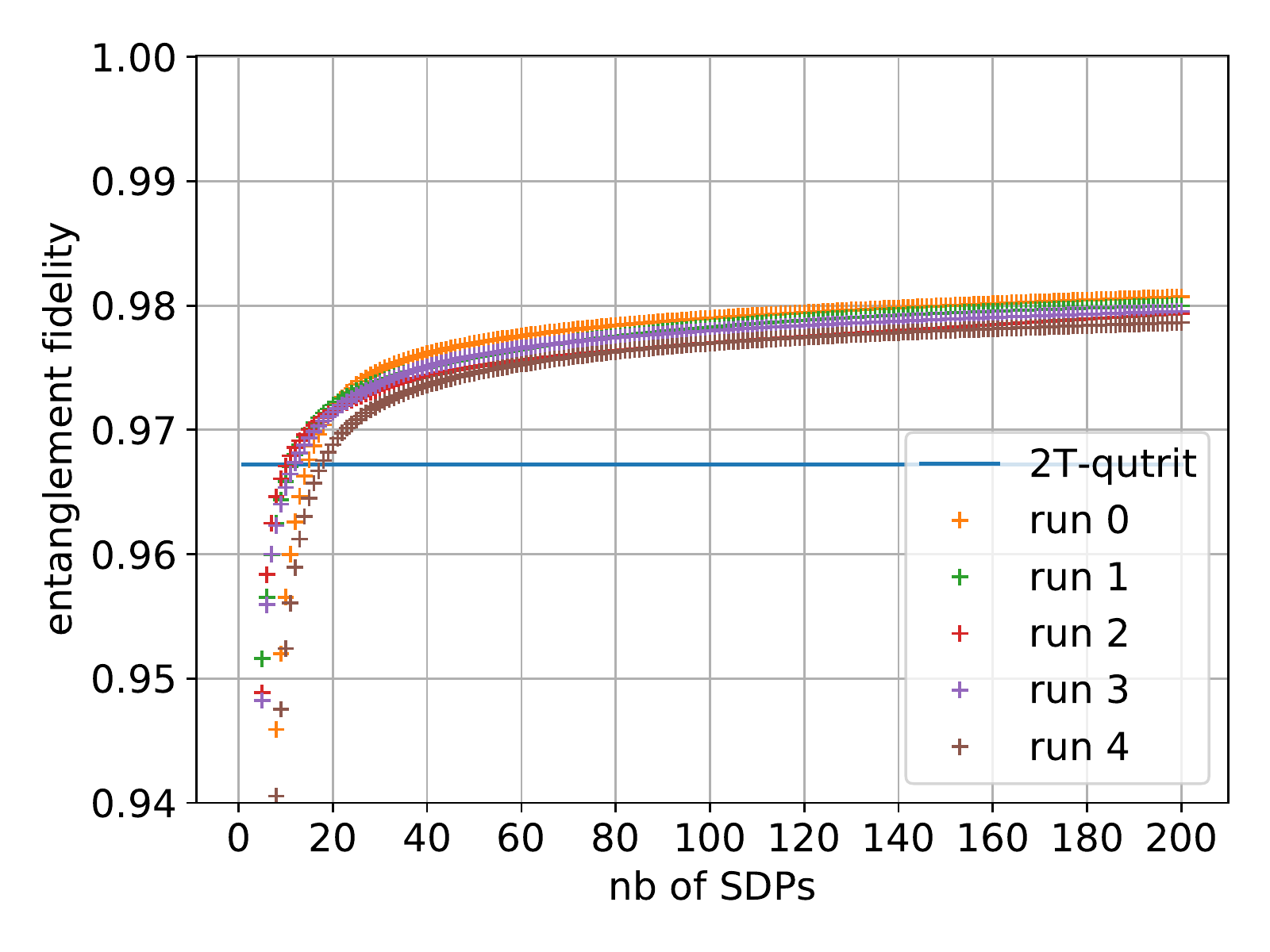}
 \includegraphics[width=0.4735\textwidth]{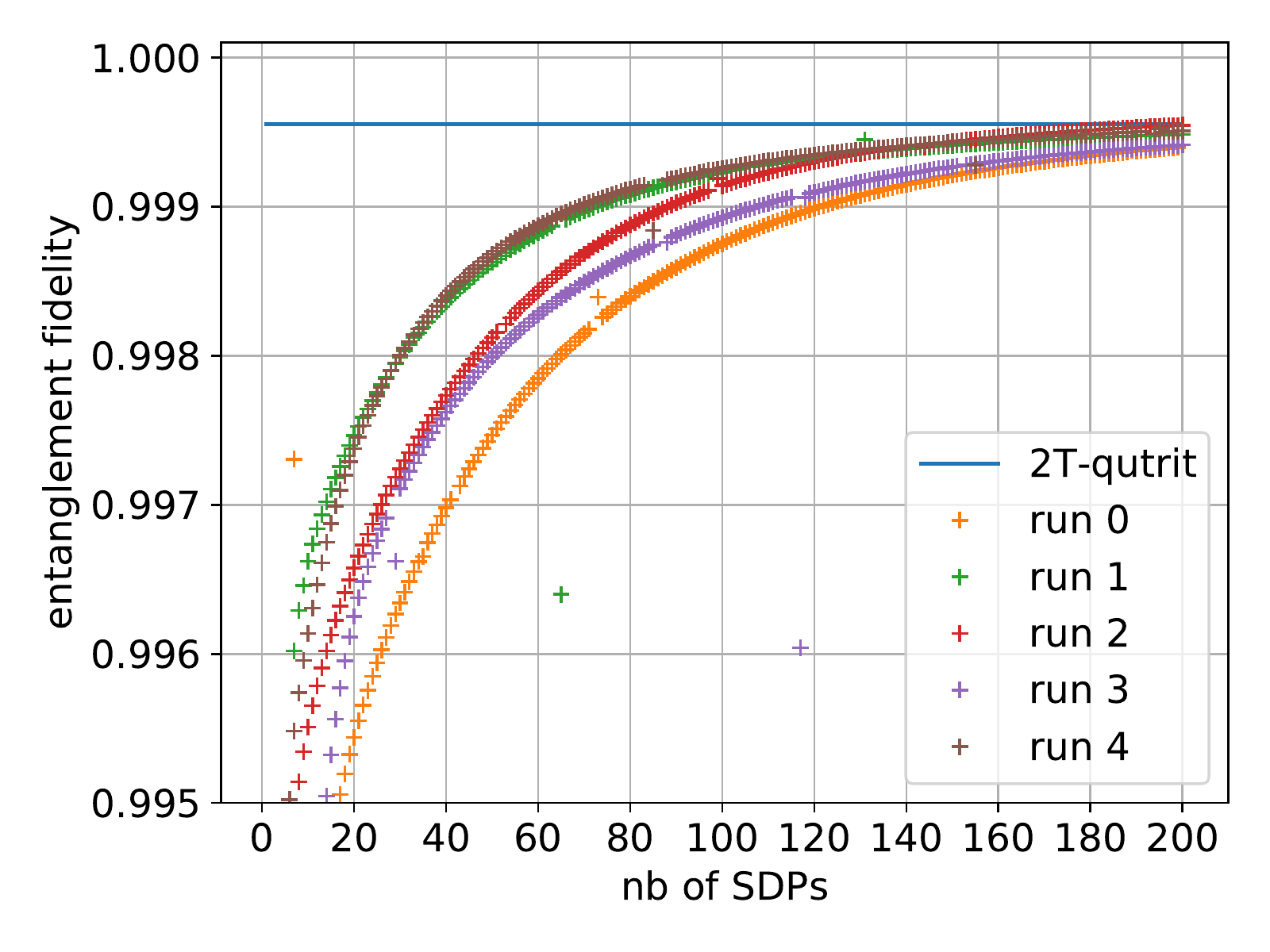}
\caption{Entanglement fidelity as a function of the number of optimisation steps, for five runs with random initial qutrit encodings in the $2T$-constellation, for $\alpha=1.5$. The fidelity for the $2T$-qutrit is also shown for comparison. Left panel: $\gamma = 0.1$, right panel: $\gamma = 0.01$.}
\label{fig:fct-sdp}
\end{figure}

We now apply the method described in Section \ref{sec:optim} to the $2T$-qutrit and compare its performance to random encodings in the $2T$-constellation. For the $2T$-qutrit, the encoding map is simply $|i\rangle \mapsto |\bar{i}\rangle$. For each encoding, we then apply the iterative optimisation procedure described above. 
In Figure \ref{fig:fct-sdp}, we plot the entanglement fidelity $F_\gamma(\mathcal{E}, \mathcal{R}):=F(\mathcal{E} \circ {\mathcal{N}}_{L,\gamma} \circ \mathcal{R})$ as a function of the number of iteration steps performed in the simulation. The comparison is done for $\alpha=1.5$, which turns out to be close to the optimal value for the $2T$-qutrit.

A first observation is that the $2T$-qutrit is indeed a fixed point of the biconvex optimisation problem and thus a local optimum.  Moreover, in the low-loss regime, the iterative optimisation procedure does not find much better encodings than the $2T$-qutrit when starting with random initial encoding (see right panel of Fig.~\ref{fig:fct-sdp}). This gives evidence that the $2T$-qutrit encoding may be close to optimal for the protection against pure loss in that regime.

\begin{figure}[htbp]
\centering
 \begin{subfigure}[b]{0.49\textwidth}
 \centering
 \includegraphics[width=\textwidth]{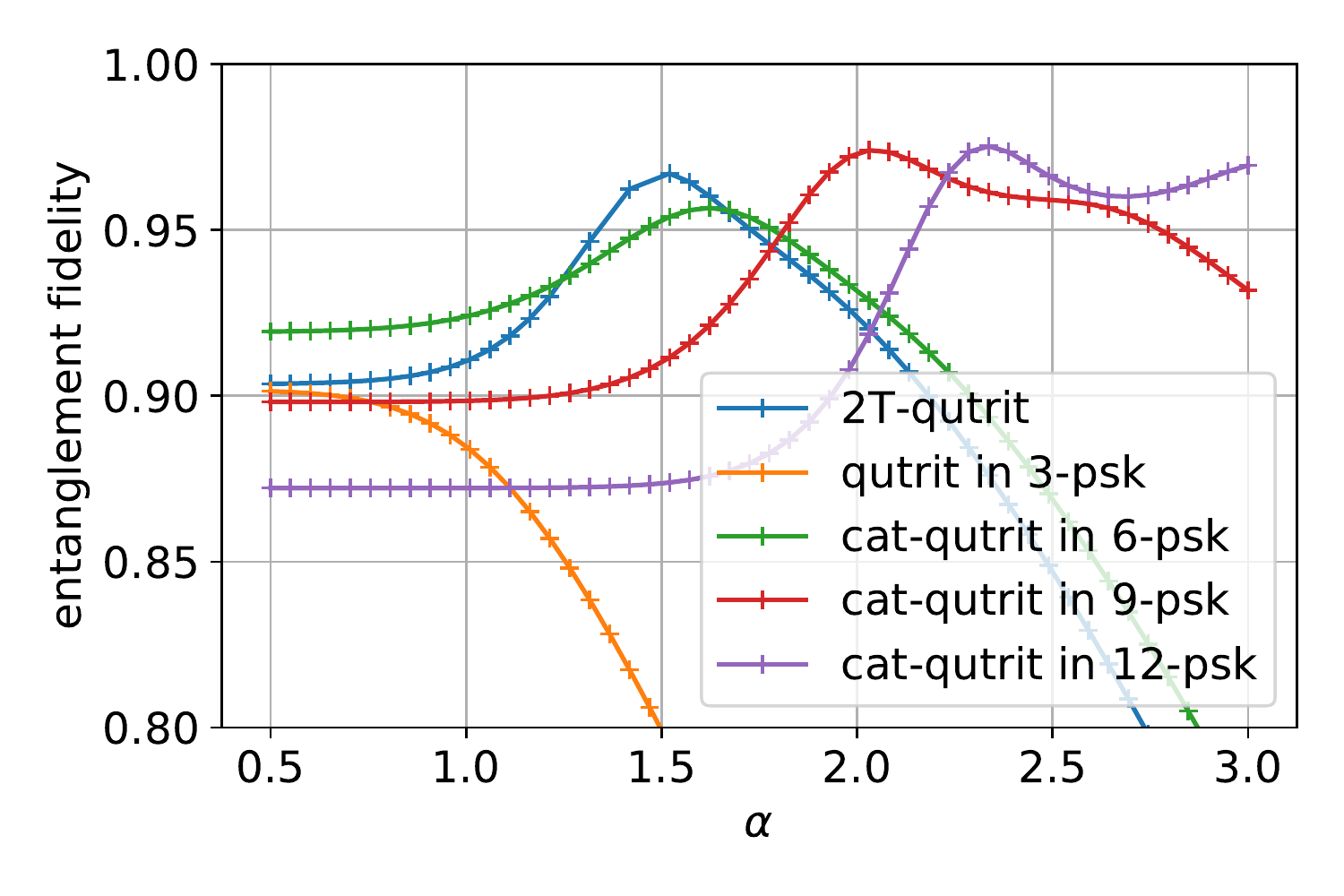}
 \caption{$\gamma = 0.1$}
 \end{subfigure}
\hfill
\begin{subfigure}[b]{0.49\textwidth}
 \centering
 \includegraphics[width=\textwidth]{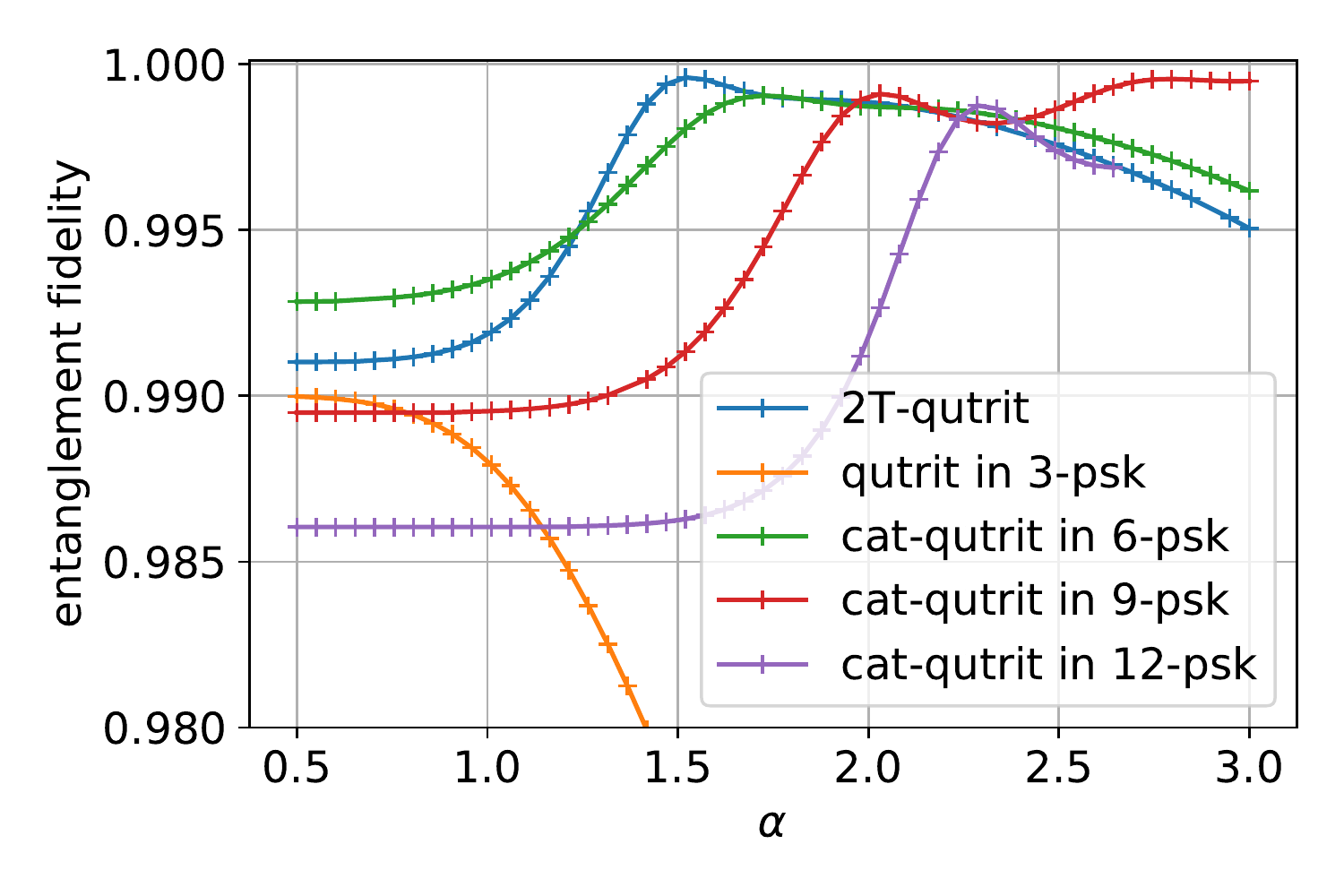}
 \caption{$\gamma=0.01$}
 \end{subfigure}
\caption{Entanglement fidelity as a function of $\alpha$, for the $2T$-qutrit and cat-qutrits with 3, 6, 9 or 12 components. Left: $\gamma = 0.1$, right: $\gamma = 0.01$.}
\label{fig:fct-nbar}
\end{figure}

It is also instructive to compare the performances of the $2T$-qutrit to that of single-mode bosonic qutrits. The cat qutrits are a family of codes that generalise the cat qubits to dimension 3. For $\ell \in \lbrace 0, 1, 2 \rbrace$, the logical states of the cat qutrit of order $3n$, are defined as superpositions of the form $\sum_{k=0}^{3n-1} \omega^{-k \ell} |\alpha \omega^{k\ell}\rangle$, where $\omega = e^{2\pi i/3n}$  and $\alpha >0$ is a free parameter.  We call this code the \emph{$3n$-PSK qutrit} since its constellation is that of a Phase-Shift Keying modulation.
In Figure \ref{fig:fct-nbar}, we compare the performance of the $2T$-qutrit to that of single-mode cat PSK qutrits, as a function of $\alpha$. 
We observe that for reasonable values of $\alpha$, the $2T$-qutrit compares favourably to single-mode encodings. One also remarks that, similarly to cat encodings, there exist optimal values (known as sweet spots) of $\alpha$ and that a larger value of $\alpha$ does not always result in a better performance.

\begin{figure}[!h]
\centering
\includegraphics[width=0.8\textwidth]{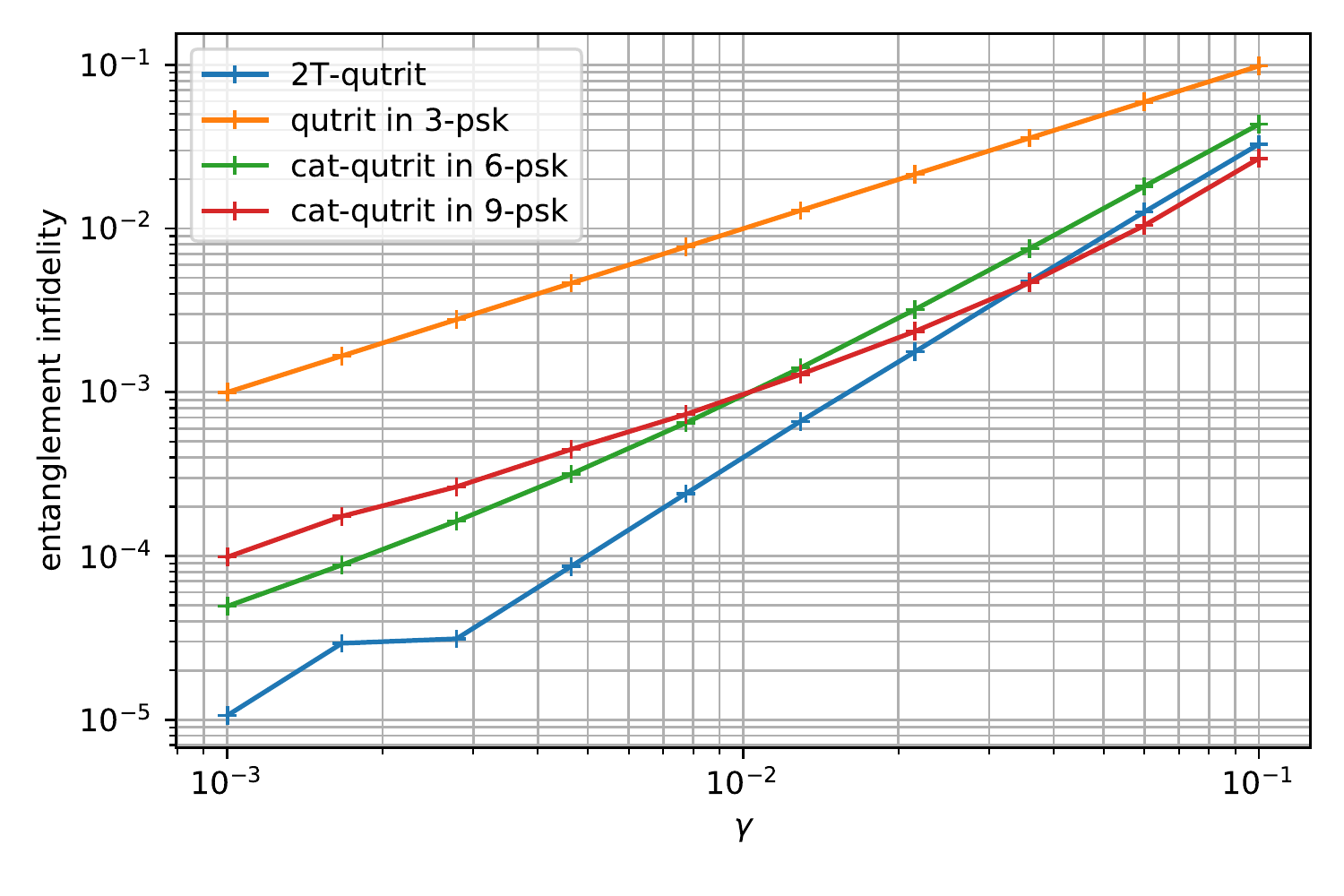}
\caption{Smallest entanglement infidelity, $1-F_\gamma$, as a function of the loss parameter $\gamma$ when $\alpha$ is optimised in the range $[ 0.25, 2]$.}
\label{fig: fct gamma}
\end{figure}

In Fig.~\ref{fig: fct gamma}, we compare the performance of the $2T$-qutrit with that of single-mode cat qutrits as a function of loss. Here the value of $\alpha$ is optimised for each bosonic code, for values in the range $[0.25, 2]$.
As already noted, we see that for reasonable values of $\alpha$, the $2T$-qutrit compares favourably to single-mode codes as soon as the loss level is sufficiently small.

\begin{figure}
    \centering
    \includegraphics{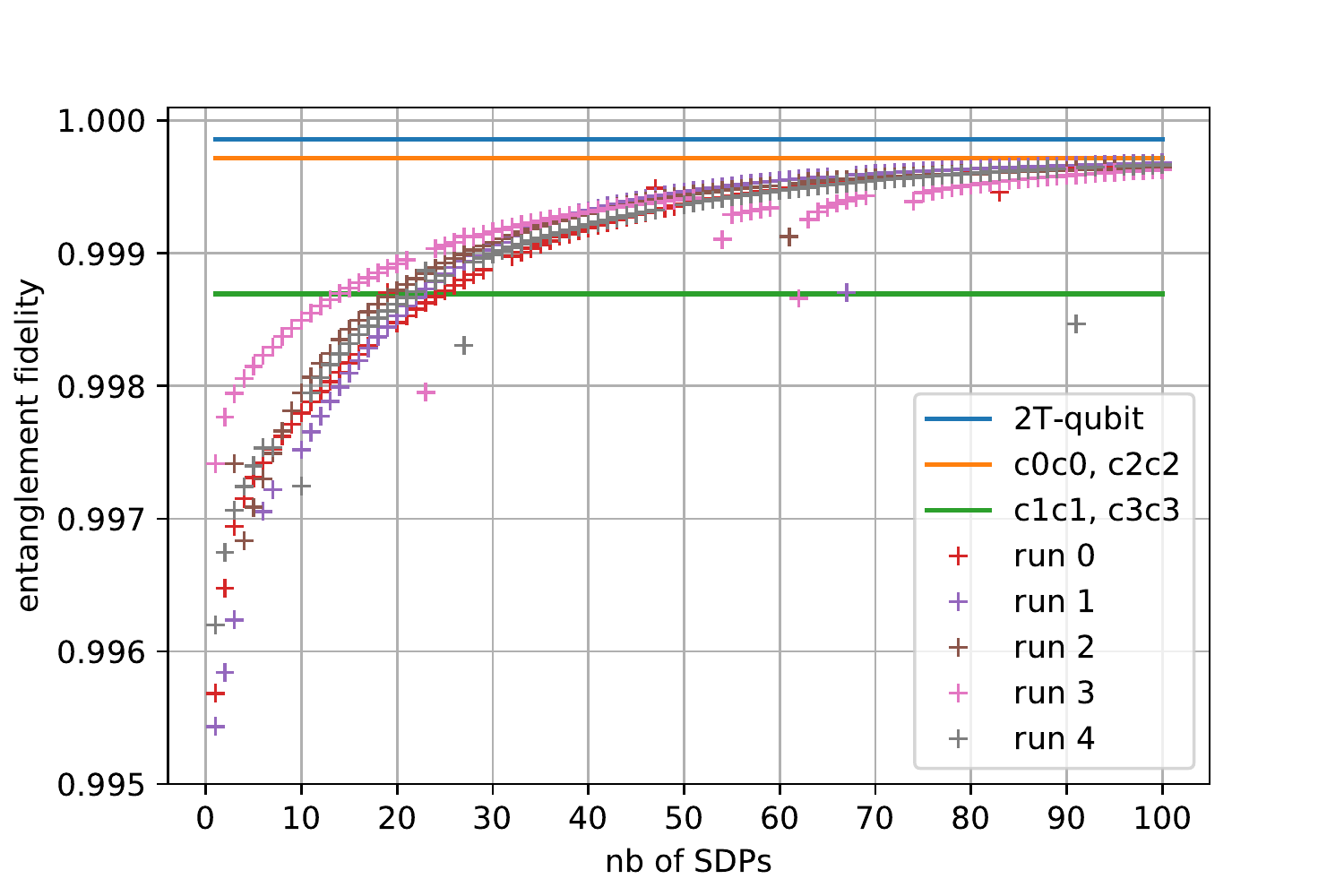}
    \caption{Entanglement fidelity of various qubit encodings, for $\gamma = 0.01$ and $\alpha=1.5$. }
    \label{fig:qubit}
\end{figure}

Finally, Fig.~\ref{fig:qubit} shows the performance of the $2T$-qubit defined in Section \ref{sec:qubit} compared to the qubits $\mathrm{Span} (\{|c_0\rangle |c_0\rangle, |c_2\rangle |c_2\rangle\})$ and $\mathrm{Span} (\{|c_1\rangle |c_1\rangle, |c_3\rangle |c_3\rangle\})$, where 
\begin{equation}
    | \mathrm{c}_0\rangle \propto \sum_{k=0}^3 |i^k \alpha\rangle, \, |\mathrm{c}_1\rangle \propto \sum_{k=0}^3 (-i)^k |i^k \alpha\rangle, \,|\mathrm{c}_2\rangle \propto \sum_{k=0}^3 (-1)^k |i^k \alpha\rangle, \, |\mathrm{c}_3\rangle\propto \sum_{k=0}^3 i^k |i^k \alpha\rangle,
\end{equation}
and to random qubit encodings in the $2T$-qutrit space.
Again, these three qubits correspond to local optima of the entanglement fidelity and that the $2T$-qutrit compares well with numerically optimised encodings in the low-loss regime.

\subsection{Performance for the dephasing channel}
\label{sub:dephasing}

While loss is certainly the major source of imperfection for many bosonic systems, it is also instructive to consider other kinds of noise, such as dephasing~\cite{LW05}.
The single-mode bosonic pure-dephasing channel is defined as
\begin{align}
\mathcal{N}_{D, \gamma} (\rho) := \sum_{m,n=0}^\infty e^{-\frac{1}{2} \gamma (m-n)^2} \langle m |\rho |n\rangle |m\rangle \langle n|,
\end{align}
where $\gamma$ now characterises the dephasing strength.
As already mentioned, we will consider two independent realisations of this channel, and therefore consider the two-mode pure-dephasing channel
\begin{align}
\mathcal{N}_{D, \gamma} \otimes \mathcal{N}_{D, \gamma} (\rho) := \sum_{\substack{m_1, m_2,\\n_1,n_2=0}}^\infty e^{-\frac{1}{2} \gamma \left((m_1-n_1)^2+(m_2-n_2)^2\right)} \langle m_1, n_1|\rho |m_2,n_2 \rangle |m_1,m_2\rangle \langle n_1,n_2|.
\end{align}

This channel admits an infinite number of Kraus operators, and it is not possible to exploit the same trick as for the pure-loss channel since a coherent state is not mapped to a pure state \textit{via} the dephasing channel. It is therefore needed to truncate the Hilbert space by keeping only the Fock states containing less than $N$ photons in total. 
We can however observe that the dephasing channel leaves invariant the photon number in each mode. This implies that to compute the entanglement fidelity of the $2T$-qutrit, it is sufficient to restrict the truncated Fock space to 
\begin{equation}
    F_{\leq N} = \mathrm{Span} \Big( |n_1, n_2\rangle \: : \: n_1 + n_2 \leq N, \quad n_1+ n_2 \equiv 0 \mod 4, \quad n_1 \equiv 0 \mod 2\Big),
\end{equation}
since the optimal recovery map will not change the photon numbers either. 
Taking $N= 4p$, we get $\mathrm{dim} \, F_{\leq N} = \frac{1}{2} (p+1)(p+2) + \frac{1}{2} p(p+1)$,
where the first term counts the pairs with $n_1 \equiv 0 \mod 4$ and the second term counts the pairs with $n_1 \equiv 2 \mod 4$. This gives
\begin{equation}
    \mathrm{dim} \, F_{\leq 4p} = (p+1)^2,
\end{equation}
which is a reduction by a factor of almost 8 compared to the naive $(4p+1)(4p+2)/2$. 

\begin{figure}[h]
\centering
 \includegraphics[width=0.7\textwidth]{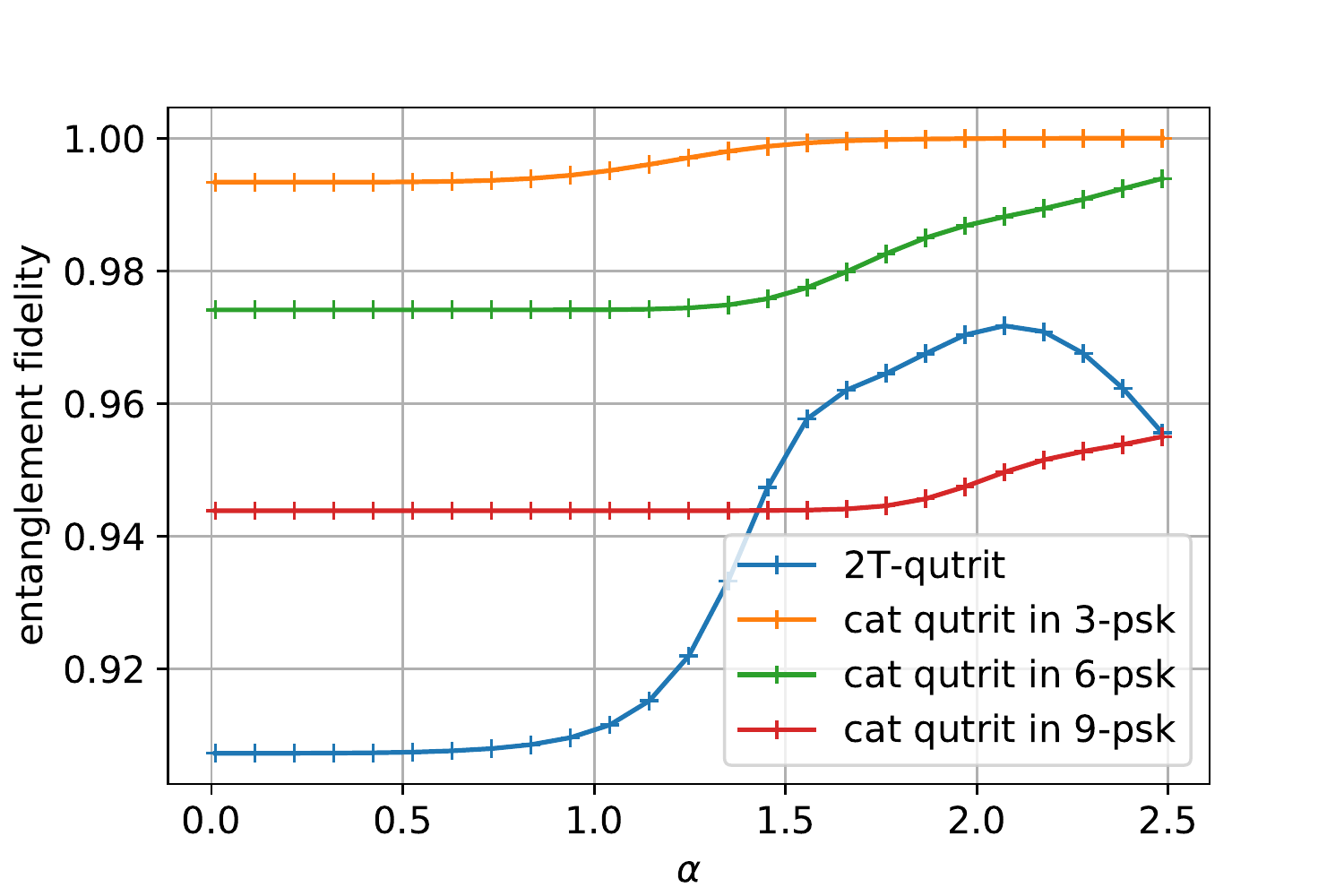}
\caption{Entanglement fidelity for the dephasing channel $\mathcal{N}_{D, \gamma}$ with $\gamma=0.01$, for the $2T$-qutrit and single-mode cat qutrits. }
\label{fig:dephasing}
\end{figure}

We plot the results on Fig.~\ref{fig:dephasing}. We first note that the tolerance to dephasing of the single-mode qutrits deteriorates quickly with the number of states in the constellation, but generally improves with increasing $\alpha$. In the regime of moderate energy, corresponding to $\alpha \leq 2.5$ here, we observe that the performance of the $2T$-qutrit presents a sweet spot, exactly as in the case of the pure-loss channel. We suspect that better fidelities could be obtained with much larger values of $\alpha$, but our simulations cannot handle this regime at the moment. 
We note that the assumption that both modes suffer independent phase noise might be too pessimistic, and that more realistic noise models are likely to be correlated, which should improve the performance of the $2T$-qutrit. 

\section{Conclusion}

In this paper, we have considered a 2-mode generalisation of bosonic cat codes. 
By working with a finite set of 24 coherent states corresponding to the finite multiplicative subgroup $2T$ of the quaternions, we can search for interesting qudits within this 24-dimensional space. 
Exploiting the decomposition of this group as the semi-direct product of the cyclic group $C_3$ with the quaternion group $Q$, we have defined the $2T$-qutrit which corresponds to a three-dimensional subspace of the 2-mode Fock space. 
Numerical simulations suggest that this bosonic qutrit may be particularly tolerant to photon-loss in the regime of low loss, at least when coupled with an ideal recovery map. 
The tolerance to dephasing is more limited, however, at least for reasonable energies. 

More importantly, it is possible to leverage the group structure of the binary tetrahedral group $2T$ to study the properties of the qutrit. In particular, we have found jump operators that can stabilise the 24-dimensional manifold corresponding to the space spanned by the 24 coherent states, and have identified a complete set of stabilisers for the $2T$-qutrit. It is also possible to define a logical $Z$-operator on the qutrit. 
Interestingly, we have finally defined a $2T$-qubit which admits a logical $X$-gate as well as a logical phase gate $P(2\pi/3)$ that can both be implemented by passive Gaussian transformations in phase space. In addition, a specific state of this qubit corresponds to two copies of a logical cat qubit state with 4 components, and recent progress in the implementation of cat qubits suggests that such states could be implemented in the near-term future. It is reasonable to expect that many techniques relevant for the preparation, manipulation and measurement of cat qubits can be ported to the setting of the $2T$-qutrit. 

We leave many open questions for future work. Probably the most intriguing one would be to understand whether it is possible to devise a universal set of gates for the $2T$-qutrit (or the $2T$-qubit). While experimental implementations will likely be very challenging, it is natural to look for multimode bosonic codes generalising the cat qubit. We have focussed here on the $2T$ subgroup of the quaternions but the binary octahedral and binary icosahedral groups are other natural candidates, at least as a purely theoretical endeavour.

\section*{Acknowledgements}
We thank Omar Fawzi for discussions about the optimisation procedure for the entanglement fidelity. We thank Victor Albert for pointing out references on group codes in the classical setting and making the link with the CLY code. 
We acknowledge support from the E.~C.~project CiViQ and from the Plan France 2030 through the project ANR-22-PETQ-0006.


\appendix

\section{Action of the pure-loss channel on finite superpositions of coherent states}
\label{sec:kraus}

The goal of this section is to find a more compact Kraus representation of the pure-loss channel of \eqref{eqn:loss} when the input state is restricted to the span of a finite number of (possibly multimode) coherent states $|\alpha_1\rangle, \ldots,| \alpha_m\rangle$. In this case, we will exhibit a representation of the channel with only $m$ Kraus operators.

First, observe that the pure-loss channel sends a coherent state $\ket{\alpha}$ onto a coherent state $\ket{\mu \alpha}$ with $\mu = \sqrt{1-\gamma}$ and therefore the output space obtained after the channel is the span of $|\mu \alpha_1\rangle, \ldots, |\mu \alpha_m\rangle$. 
It is useful to consider orthonormal bases of both spaces. Let us denote by $\tau$ and $\tau'$ the uniform mixtures of the coherent states in the input and output spaces:
\begin{align}
\tau := \frac{1}{m} \sum_{k=1}^{m} \ket{\alpha_k}\bra{\alpha_k}, \qquad 
\tau' := \frac{1}{m} \sum_{k=1}^{m} \ket{\mu\alpha_k}\bra{\mu\alpha_k}.
\end{align}
One can check that the sets $\{ |\psi_k\rangle\}_{k \in [m]} $ and $\{ |\psi'_k\rangle\}_{k \in [m]} $ form orthonormal bases of $\mathrm{Span}(\{|\alpha_1\rangle, \ldots, |\alpha_m\rangle\})$ and $\mathrm{Span}(\{|\alpha'_1\rangle, \ldots, |\alpha'_m\rangle\})$ respectively, with 
\begin{align}
|\psi_k\rangle := \frac{1}{\sqrt{m}} \tau^{-1/2} \ket{\alpha_k}, \qquad |\psi_k'\rangle := \frac{1}{\sqrt{m}} \tau'^{-1/2} \ket{\mu \alpha_k}.
\end{align}
To see this, observe that
\begin{align}
\sum_{k=1}^m |\psi_k\rangle \langle \psi_k| = \frac{1}{m} \tau^{-1/2} \sum_{k=1}^m|\alpha_k\rangle \langle \alpha_k| \tau^{-1/2} = \tau^{-1/2} \tau \tau^{-1/2}
\end{align}
which is the projector onto $\mathrm{Span}(\{|\alpha_1\rangle, \ldots, |\alpha_m\rangle\})$.

Let us define the operators
\begin{align}\label{eqn:good}
  C_k := \sum_{\ell=1}^m \langle \psi'_k|\sqrt{\gamma} \alpha_\ell\rangle {\tau'}^{1/2} |\psi'_\ell \rangle \langle \psi_\ell|\tau^{-1/2}
\end{align}
for $k \in [m]$.
We claim that they form a set of Kraus operators for the pure-loss channel acting on $\mathrm{Span}(\{ |\alpha_1\rangle, \ldots, |\alpha_m\rangle\})$. We check this by computing their action on $|\alpha_i\rangle \langle \alpha_j|$ for arbitrary $i,j \in [m]$:
\begin{align}
\sum_{k=1}^m C_k |\alpha_i\rangle \langle \alpha_j| C_k^\dag &= m \sum_{k=1}^m C_k \tau^{1/2} |\psi_i\rangle \langle \psi_j| \tau^{1/2} C_k^\dag\\
&= m \sum_{k=1}^m \langle \psi'_k | \sqrt{\gamma} \alpha_i\rangle \tau'^{1/2} |\psi_i'\rangle \langle \psi_j'| \tau'^{1/2} \langle \sqrt{\gamma} \alpha_j | \psi'_k\rangle\\
&= m  \langle \sqrt{\gamma} \alpha_j | \Big(\sum_{k=1}^m |\psi'_k\rangle \langle \psi'_k | \Big) |\sqrt{\gamma} \alpha_i\rangle \tau'^{1/2} |\psi_i'\rangle \langle \psi_j'| \tau'^{1/2}\\
&=  \langle \sqrt{\gamma} \alpha_j | \sqrt{\gamma} \alpha_i\rangle |\mu \alpha_i\rangle \langle \mu \alpha_j|\\
&= \mathcal{N}_{L,\gamma} (|\alpha_i\rangle \langle \alpha_j|).
\end{align}
Moreover, the operators are correctly normalised since 
\begin{align}
\sum_{k=1}^m C_k^\dag C_k&= \sum_{i,j,k=1}^m \langle \sqrt{\gamma} \alpha_i | \psi'_k \rangle\langle \psi'_k|\sqrt{\gamma} \alpha_j \rangle {\tau}^{-1/2} |\psi_i \rangle \langle \psi'_i|\tau'^{1/2} \tau'^{1/2} |\psi'_j\rangle \langle \psi_j| \tau^{-1/2}\\
&= \frac{1}{m^2}\sum_{i,j=1}^m \langle \sqrt{\gamma} \alpha_i | \sqrt{\gamma} \alpha_j \rangle {\tau}^{-1} |\alpha_i \rangle \langle \mu \alpha_i |\mu\alpha_j\rangle \langle \alpha_j| \tau^{-1} \\
&= \tau^{-1} \Big( \frac{ 1 }{m^2} \sum_{i,j=1}^m \langle \alpha_i | \alpha_j \rangle |\alpha_i \rangle \langle \alpha_j| \Big)\tau^{-1} = \tau^{-1} \tau^2 \tau^{-1}
\end{align}
which is the projector onto the input space.

\section{Orthonormal basis of the $2T$-qutrit}
\label{app:B}

All the relevant states studied in this work appear as superpositions of coherent states and it is convenient to gather here the values of the various normalisation constants and overlaps. 

First, the overlap between two single-mode coherent states $|\alpha\rangle$ and $|\beta\rangle$ is 
\begin{align}
\langle \alpha | \beta\rangle = e^{-\frac{|\alpha|^2}{2} - \frac{|\beta|^2}{2} + \alpha^* \beta}.
\end{align}
Let us define the single-mode cat code states:
\begin{align}
|\alpha_2\rangle & := c_2^\alpha (|\alpha\rangle + |-\alpha\rangle),\\
|i\alpha_2\rangle & := c_2^\alpha (|i\alpha\rangle + |-i\alpha\rangle),\\
|\alpha_4\rangle & := c_4^\alpha  ( |\alpha\rangle + |i\alpha\rangle + |-\alpha\rangle + |-i\alpha\rangle),
\end{align}
with normalisation coefficients given by
\begin{align}
c_2^\alpha = \frac{1}{\sqrt{2(1+e^{-2|\alpha|^2})}}, \qquad
c_4^\alpha = \frac{1}{\sqrt{8 e^{-|\alpha|^2} (\cosh |\alpha|^2 + \cos|\alpha|^2)}}.
\end{align}
In particular, we have 
\begin{align}
|\widetilde{\phi}_0\rangle &= \frac{1}{c_4^\beta} (|\beta_4\rangle | 0\rangle + |0\rangle |\beta_4\rangle),\\
|\widetilde{\phi}_1\rangle &= \frac{1}{(c_2^\alpha)^2} (|\alpha_2\rangle |i \alpha_2\rangle + |i \alpha_2\rangle |\alpha_2\rangle),\\
|\widetilde{\phi}_2\rangle &= \frac{1}{(c_2^\alpha)^2} ( |\alpha_2\rangle |\alpha_2\rangle + |i \alpha_2\rangle |i \alpha_2\rangle),
\end{align}
with $\beta =  \alpha (1+i)$, and therefore
\begin{align}
\langle \widetilde{\phi}_0|\widetilde{\phi}_0\rangle &= \frac{1}{(c_4^\beta)^2} ( 2 + 2 |\langle 0 |\beta_4\rangle|^2 )\\
&=  \frac{2}{(c_4^\beta)^2} (1 + 16 (c_4^\beta)^2 e^{-|\beta|^2}) \\
&= \frac{2}{(c_4^\beta)^2}+ 32 e^{-2|\alpha|^2} \\
&= 16 e^{-2|\alpha|^2} (2+ \cosh 2|\alpha|^2 + \cos 2|\alpha|^2).
\end{align}
We assume that the parameter $\alpha$ is real throughout, so we get the following expression for the normalisation constant:
\begin{align}
|\phi_\ell\rangle = \nu \sum_{q\in \omega^\ell Q} |q\rangle, \quad \text{with} \quad \nu =  \frac{e^{\alpha^2} }{4 \sqrt{ \cosh(2\alpha^2)+2 + \cos(2\alpha^2)}}.
\end{align}
We note that the overlap $\langle \phi_k|\phi_\ell\rangle = \langle \phi_0 |\mathcal{U}^{\ell-k} |\phi_0\rangle$ only depends on $\ell-k$. Since $\langle \phi_0|\phi_0\rangle = 1$, we only need to compute one other overlap, say $\langle \phi_1|\phi_2\rangle$ (since $\langle \phi_2|\phi_1\rangle$ is its complex conjugate):
\begin{align}
\langle \phi_1|\phi_2 \rangle &= \nu^2 \langle \widetilde{\phi}_1|\widetilde{\phi}_2\rangle = \frac{\nu^2}{(c_2^\alpha)^4} (4 \mathrm{Re} (\langle i \alpha_2|\alpha_2\rangle)).
\end{align} 
The overlap $\langle i \alpha_2|\alpha_2\rangle$ is easily computed:
\begin{align}
\langle i \alpha_2|\alpha_2\rangle &=   (c_2^\alpha)^2 (\langle i\alpha | + \langle -i \alpha|) (|\alpha \rangle + |-\alpha \rangle)\\
&=   (c_2^\alpha)^2(\langle i\alpha |\alpha\rangle  + \langle i \alpha|-\alpha \rangle + \langle - i\alpha |\alpha\rangle  + \langle - i \alpha|-\alpha \rangle) \\
&=   4 (c_2^\alpha)^2 e^{-\alpha^2} \cos \alpha^2.
\end{align}
Injecting this in the previous expression, we obtain
\begin{align}
\langle \phi_1|\phi_2 \rangle &=16 \frac{\nu^2}{(c_2^\alpha)^2} e^{-\alpha^2} \cos \alpha^2\\
&= \frac{2 e^{\alpha^2}\cos \alpha^2 (1+e^{-2|\alpha|^2})}{ \cosh(2\alpha^2)+2 + \cos(2\alpha^2)} 
\end{align}
and finally
\begin{align}
\langle \phi_{\ell}|\phi_{\ell+1}\rangle =   \frac{4 \cosh \alpha^2 \cos \alpha^2  }{2 + \cos(2\alpha^2) + \cosh(2\alpha^2)}.
\end{align}
To find an orthonormal basis of the $2T$-qutrit, it is natural to consider states of the form
\begin{align}\label{eqn:bark}
|\bar{k} \rangle &= \nu_k \sum_{\ell=0}^2 \zeta^{-k \ell} |\phi_\ell\rangle.
\end{align}
Indeed, they are orthogonal since they correspond to an eigenbasis of the operator $\mathcal{U}$:
\begin{align}
\mathcal{U} |\bar{k}\rangle &= \nu_k \sum_{\ell=0}^2 \zeta^{-k \ell} \mathcal{U} |\phi_\ell\rangle\\
&= \nu_k \sum_{\ell=0}^2 \zeta^{-k \ell} |\phi_{\ell+1}\rangle\\
&= \zeta^k |\bar{k}\rangle.
\end{align}
Let us compute the normalisation coefficient $\nu_k$: \eqref{eqn:bark} gives
\begin{align}
\frac{1}{(\nu_k)^2}  &= \sum_{p,q=0}^2 \zeta^{k(p-q)} \langle \phi_p|\phi_q\rangle \\ 
&= 3 (1 + (\zeta^k + \zeta^{2k}) \langle \phi_1|\phi_2\rangle) 
\end{align}
and therefore
\begin{align} 
\nu_0 &= \frac{1}{\sqrt{3(1+ 2  \langle \phi_1|\phi_2\rangle)}}, \label{eqn:nu0}\\
\nu_1 =\nu_2  &= \frac{1}{\sqrt{3(1-  \langle \phi_1|\phi_2\rangle)}}. \label{eqn:nu12}
\end{align}

It is also easy to write $|\phi_k\rangle$ in the logical basis:
\begin{align}\label{eqn:phi_k}
 |\phi_k\rangle = \frac{1}{3} \sum_{\ell=0}^2 \frac{\zeta^{k\ell}}{\nu_\ell} |\bar{\ell}\rangle.
\end{align}



\end{document}